# Evaluation scheme for the design of high power single mode vertical-cavity surface-emitting lasers

**Young-Gu Ju**

*KyungPook National University, Sangyeok, Daegu 702-701 Korea*
ygju@knu.ac.kr

**Abstract:** A very simple and efficient evaluation procedure is suggested for the design of high power single mode VCSELs by reviewing the physical mechanisms that governs mode transition and simplifying the computation steps. In addition, the new structures are proposed and tested following the suggested evaluation procedure. As a result, the proposed design exhibits much better stability of the fundamental mode over a current range wider than the conventional one.

---

### References and links

1. C. Jung, R. Jager, M. Grabherr, P. Schweitzer, R. Michalzik, B. Weigl, S. Muller, and K. J. Ebeling, "4.8 mW single mode oxide confined top surface emitting vertical cavity laser diodes," Electron. Lett. **33**, 1790-1791 (1997).
2. H. Unold, S. W. Mahmoud, F. Eberhard, R. Jaeger, M. Kicherer, F. Mederer, M. C. Riedl, and K. J. Ebeling, "Large-area single-mode selectively oxidized VCSELs: Approaches and Experimental," Proc. SPIE, **3946**, 207-218 (2000).
3. K. D. Choquette, A. A. Allerman, K. M. Geib, and J. J. Hindi, "Lithographically-defined gain apertures within selectively oxidized VCSELs," in CLEO Tech. Dig. 2000, 232-233 (2000),
4. Y. A. Wu, G. S. Li, R. F. Nabiev, K. D. Choquette, C. Caneau, and C. J. Chang-Hasnain, "Single-mode, passive antiguide vertical cavity surface emitting laser," IEEE J. Select. Topics Quantum Electron. **1**, 629-637 (1995)
5. H. Martinsson, J. A. Vukusic, and A. Larsson, "Single-mode power dependence on surface relief size for mode-stabilized oxide-confined vertical-cavity surface-emitting lasers," IEEE Photon. Technol. Lett. **12**, 1129-1131 (2000).
6. J. A. Vukusic, H. Martinsson, J. S. Gustavsson, and A. Larsson, "Numerical optimization of the single fundamental mode output from a surface modified vertical-cavity surface-emitting laser", IEEE J. of Quantum Electron., **37**, 108-117 (2001)
7. D. I. Babic, G. H. Gohler, J. E. Bowers, and E. L. Hu, "Isotype heterojunctions with flat valence or conduction band," IEEE J. Quantum Electron. **33,** 2195-2198 (1997).
8. E. R. Hegblom, "Engineering oxide apertures in vertical cavity lasers," Ph.D dissertation, Electrical and Computer Engineering, University of California (1999).
9. Takanori Okoshi, *Optical fibers*, (Academic Press 1982), Chap. 4.
10. Y. G. Ju, J. H. Ser, and Y. H. Lee, "Analysis of metal-interlaced-grating vertical-cavity surface-emitting lasers using the modal method by modal expansion", IEEE J. of Quantum Electron., **33**, 589-595 (1997)
11. G. R. Hadley, "Effective index model for vertical-cavity surface-emitting lasers," Optics Lett. **20**, 1483-1485 (1995)
12. W. H. Press, S. A. Teukolsky, W. T. Vetterling, B. P. Flannery, *Numerical recipes in C* (Cambridge University Press, 1982), Chap. 17.

---

## 1. Introduction

Vertical-cavity surface-emitting lasers(VCSELs) have gained great attention as an attractive light source for short-distance optical communication systems, optical interconnects, optical storage, and laser printing. Many applications of VCSELs, however, are still limited by the difficulty of high power fundamental mode operation in standard devices. Due to the extremely short cavity length, the VCSEL has only one longitudinal mode and the large transverse dimension leads to lasing in multiple transverse modes. There have been numerous approaches in trying to solve this issue. They include small aperture oxide VCSELs[1], an extended optical cavity[2], gain apertured VCSELs[3], anti-guided structures[4] and the use of surface relief structures[5]. Some of the results are excellent, but some problems remain in the single mode power level, single mode operating current range, device-reliability and fabrication complexity. Along with the experimental results, many theoretical models are also suggested and several of them are in good correlation with the experimental results. The excessive effort required to achieve a high accuracy makes the modeling so complicated and often prevents providing physical insights about the key mechanism of mode control in VCSELs to the device engineers[6]. In this paper, the importance of the thermal issue in mode control is emphasized and a simple formula describing the relationship of the device temperature to device geometry is provided. Focusing on this thermal problem, a new structure is proposed in order to improve mode stability over a wider operating current range and it is compared with the conventional structure through simple calculations.

## 2. Size dependence of temperature

In order to achieve a high power fundamental mode, there is a need to remember what causes output power to saturate in VCSELs. The power rollover can be mainly attributed to the mismatch between gain peak and cavity peak as the device temperature increases, especially in the case of a short wavelength. Since a small active volume means a high electrical and thermal resistance, a small device has a higher temperature increase for a given current and thus has earlier output power saturation. Therefore, if the device is large enough to ignore optical scattering loss, the best way to achieve high power operation is to suppress the temperature rise in the device. The temperature rise inside the cavity is a function of electrical resistance and thermal conductance of the device since they correspond to heat generation and dissipation. Therefore, historically, a lot of effort has been paid in reducing electrical resistance by modifying the doping profile around a hetero-junction barrier in the top distributed Bragg reflector (DBR)[7]. The trials proved to be successful and contributed to commercialization of 850 nm VCSELs. Despite the great success in reducing electrical resistance, the investigation on the relation between geometry and temperature profile revealed a fundamental limit of heat dissipation efficiency in correlation with size dependence.

Table 1. Thermal conductivity used in calculation

| Material | K(W/(m K)) |
|---|---|
| GaAs | 44. |
| AlGaAs | 16. |
| Au | 3150. |

At first, the electrical resistance is inversely proportional to the square of the radius of the aperture in a pillar structure in accordance with Ohm’s law. However, if the current spreads well above the aperture in place with a large contact pad, it is approximately proportional to 1/r [8]. Second, the thermal resistance also shows a similar dependence on the radius of the current aperture since both temperature and electrical potential follow Laplace’s equation with the exception of the source term. For a more exact expression of geometrical thermal dependence, the numerical simulation was performed using a commercial FEM(Finite Element Method) solver called ANSYS. In the calculation, the geometry of a standard 850 nm oxide VCSEL is used and the core temperatures are measured for the same amount of heat generated. The parameters used in the calculation are shown in Table 1. Since the thermal conductance of the composing materials is relatively well known and the equation solving software has been proven successful in mechanical engineering, the result is expected to agree with real values. For instance, the core temperature with a radius of 5 μm rises 11.5 °C for heat dissipation of 7.8 mW. This temperature increase is within the nominal operating temperature range, which is usually measured by observing the wavelength shift. The final result is summarized in Fig. 1. The log-log plot indicates that the temperature increase is roughly proportional to $1/r^{1.1}$. After all, the core temperature rise is proportional to $1/r^{2.1}$ in oxide type or $1/r^{3.1}$ in an etched pillar type for a fixed electrical current. It comes from the fact that the heat generation is proportional to electrical power dissipation and the electrical power dissipation is again proportional to electrical resistance for a constant current. The strong size dependence of the device temperature puts a lower boundary on the size of the current aperture because a higher temperature causes early power saturation and limits the optical peak power.

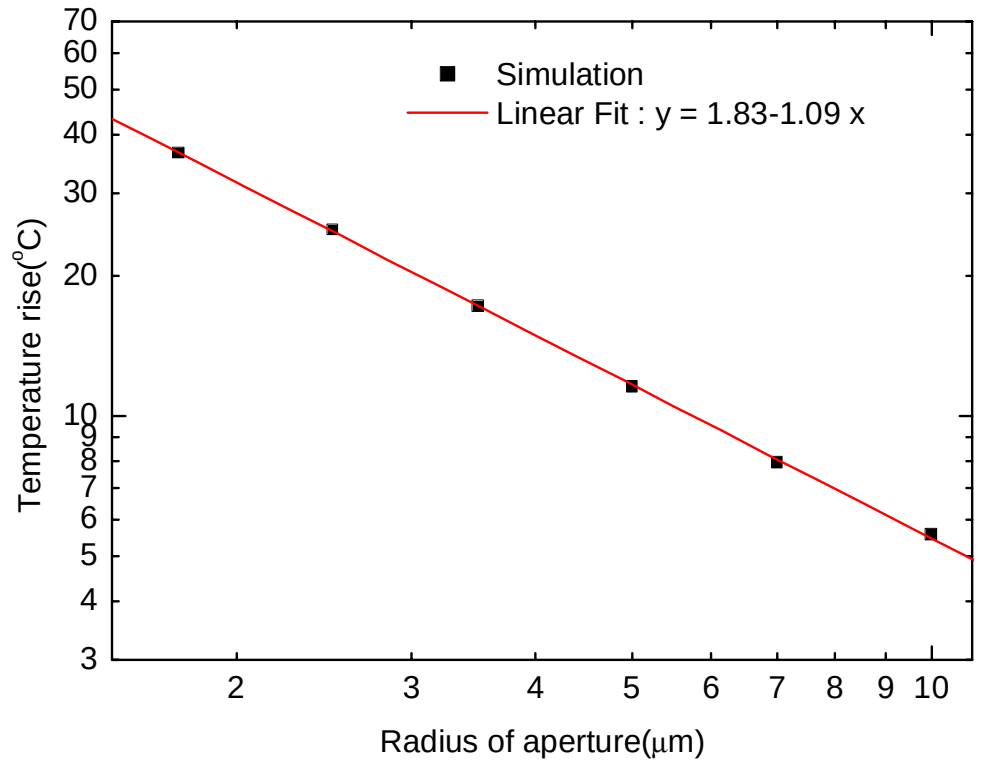

Fig. 1. Temperature rise versus the radius of the current aperture.

Actually, the size dependence of the temperature limits not only the design of high power VCSELs, but also the design of single mode VCSELs because the temperature profile severely influences transverse mode behavior especially at small aperture. Reviewing the thermal lens behavior in an implant VCSEL helps to understand the effect of a thermal profile on the mode transition of various VCSEL structures. The current path produces a heat profile similar to its own due to the relation between power dissipation and heat profile. Thus, the size of the current aperture can be assumed to be equal to that of the heat source. Heat generation leads to a temperature profile based on a heat conduction equation. The temperature profile changes the refractive index of the cavity. Finally, the central region has the increased effective index profile acting like a lens. This phenomenon implies that we cannot separate the current path and refractive index change and the index profile changes as the current increases. Even if a VCSEL starts with a single mode, the increased index around the current path makes higher order modes emerge as the current increases. This trend gets worse with a larger diameter aperture because the single mode condition depends on the product of the refractive index difference and the diameter of core[9]. It is thought that the small current aperture may maintain the fundamental mode over wide current range. In reality, its temperature increases rapidly with a smaller aperture according to the dependence on size of the temperature rise. For a small current aperture, the higher index step prevents a single mode operation or the increased core temperature of the device stops obtaining high output power. Therefore, it is obvious how important it is to investigate modal behavior as a function of the temperature increase at the center in order to find the optimum design for a single mode VCSEL.

### 3. Evaluation procedure

The above-mentioned viewpoint can be used to establish an efficient evaluation procedure for a new design aiming at high power single mode VCSELs. In the evaluation procedure, the mode behavior is usually described by modal gain or modal loss because the loss difference or gain difference is proportional to the mode suppression ratio[10]. In addition, the temperature rise can be replaced by the refractive index change of the cavity since the temperature rise is almost proportional to the index increase. Then, the evaluation procedure of mode behavior can be reduced into a plot of modal loss/gain versus the index change within the current aperture. The calculation becomes purely optical and a very simple problem. With this approach, the only concern is how large a difference the modal loss/gain maintains as the index at the center increases. The suggested procedure eliminates the need to obtain complete light-current curves by way of a complicated calculation, only to assess the performance of a new design of a single mode high power VCSEL. Moreover, the calculation of a mode profile for a given index profile is greatly simplified by the introduction of the effective index model for VCSEL[11]. This model removes the rapid vertical variation of the field from a three-dimensional field profile. In the case of cylindrical symmetry, solving a mode profile is reduced into a one-dimensional problem. Therefore, employing the effective index model for the evaluation procedure saves a lot of programming effort and computation time further. In this paper, the one-dimensional scalar wave equation, Eq. (1) is solved by a shooting algorithm[12] and the overlap integral is carried out to obtain the modal gain or loss.

$$\frac{1}{r}\frac{d}{dr}(r\frac{dR}{dr})+[\omega^2\varepsilon(r)\mu_0-\beta^2-\frac{m^2}{r^2}]R=0. \tag{1}$$

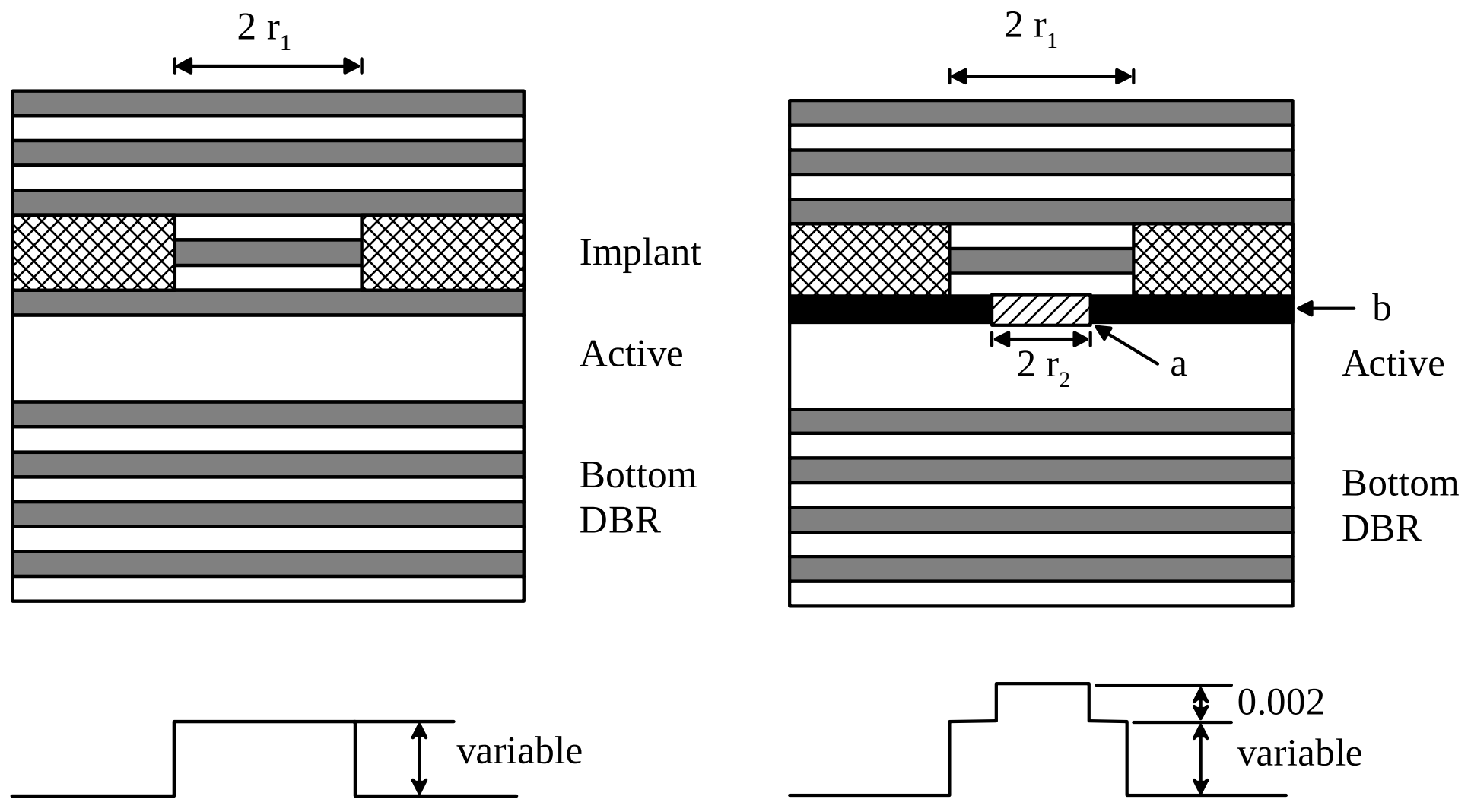

(a) Structure A (b) Structure B

Fig. 2. Two designs for high power single mode VCSELs. The plots at the bottom represent the refractive index profiles of the respective structures. Parts a and b in (b) are an index guiding disc and absorption region.

The two designs shown in Fig. 2 are tested by the suggested evaluation procedure. The first design A is selected as a control group in order to compare it to the newly proposed structure B. The calculations from the two different 850 nm VCSEL designs are presented in Fig. 3 and Fig. 4. The first structure A can be made by an implantation with the radius of the current aperture equal to 3 μm. As the refractive index at the center increases, the higher modes appear. The modal gain of the fundamental mode remains higher than those of other modes over the entire index change. The modal gain difference, however, is small in absolute value and decreases with temperature rise. In the plot, the modal gain of 0.01 can compensate the cavity loss when the reflectivity of the DBR is 0.99. Owing to the absence of loss difference in the first structure, the mode selection is mainly determined by gain difference. As expected, the number of modes is less than that of the larger current aperture. If the aperture size decreases further, the higher modes appear at a larger index change and the gain difference becomes larger. In this case, the gain difference at the index change of 0.01 is 0.0004. The mode suppression ratio with the loss difference is about 80. The mode suppression ratio heavily, however, depends on many other parameters like output power level and extra loss terms. In this paper, the value is used for comparison purposes between the structures. Even though this gain difference is small, it is still greater than that of the large current aperture VCSEL.

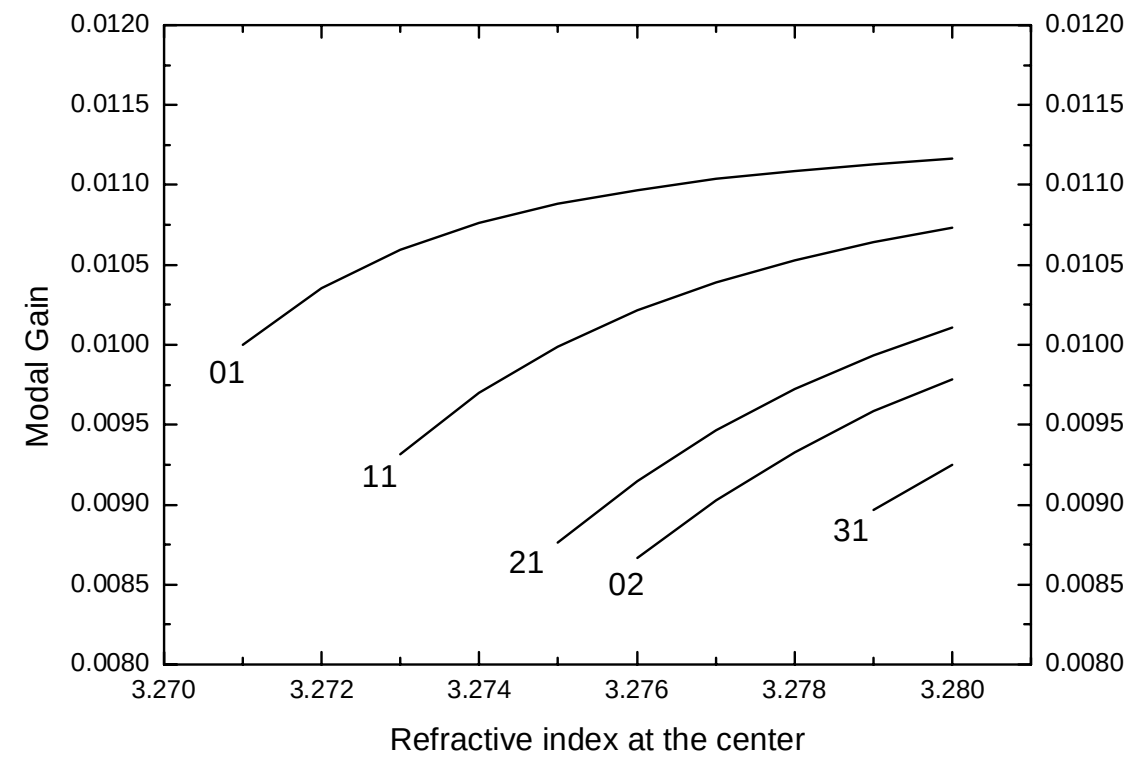

Fig. 3. Modal gains of structure A as a function of the index change at the center.

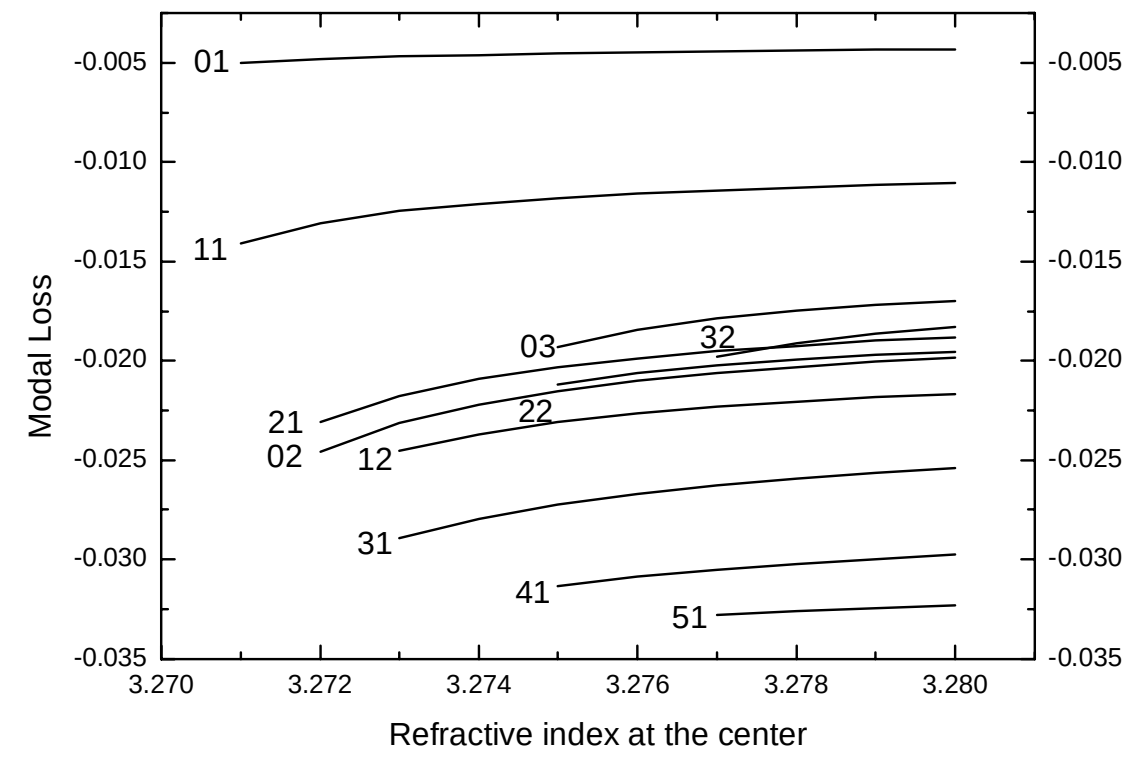

Fig. 4. Modal losses of structure B as a function of index change at the center..

From another simulation with the same structure as that of A and with an aperture of 5 µm in radius, the gain difference is only 0.0001. The small gain difference at a large index change means that the mode discrimination is very weak at a high temperature or high current. At a high current, the spatial hole burning effect often changes the modal gains and causes a mode transition by making a higher order mode have a greater modal gain than the fundamental mode. When the performance of the two current apertures is compared, the range of index change should be considered carefully. Since the temperature scale depends on $1/r^{2.1}$, the radius of 3 μm has about three times a larger temperature increase for a given current. In other words, the 3 μm device has three times smaller input current at an index change of 0.01 than the 5 μm device. In this sense, the design should be directed to increase modal loss/gain difference by sustaining a large current aperture. Before entering the second structure, more should be said regarding the horizontal axis of the plot in order to help grasp the idea how large it is in the real device. The index scale of 0.01 corresponds to a 2.6 nm shift in the cavity wavelength according to Hadley's results[11]. Considering that the cavity peak moves at the rate of 0.05 nm/°C, the temperature change is 52 °C. The elevated temperature also causes a mismatch between the cavity peak and gain peak. The offset between the two peaks is about 13 nm when temperature increases 52 °C. This mismatch can result in significant gain loss and eventually a power saturation because the gain width is around 20 nm.

In an effort to improve transverse mode behavior, the second structure B in Fig. 2 is proposed and the results of the calculation are illustrated in Fig. 4. It maintains a large current aperture and provides a selective loss mechanism to achieve a high mode suppression ratio. The radius of the current aperture is 5 μm. The structure consists of the selective absorption layer, which can be patterned by a conventional lithographic technique. The current path is defined by implantation to avoid an excessive index guiding effect. While index guiding reduces optical scattering loss, it deteriorates single mode behavior. Therefore, there should be a trade-off in design with regard to index guiding. Structure B contains an index-guiding disc causing an index difference of 0.002 at the center for a stable mode behavior. This small index step is soon buried in the large index step generated by a thermal profile as the operating current increases. The index disc, however, contributes to improving laser performance and mode stabilization right above the threshold because, without it, the VCSELs act like a conventional implant lasers with large threshold current. The modal loss difference at the index change of 0.01 is about 0.007. It is almost 18 times larger than that of A and is 70 times larger than that of the 5 μm device without selective loss structure. The increase in the modal loss difference results in the same amount of increase in the mode suppression ratio. Since many factors come into play at a high current range, the real device may hardly show such significant improvement in mode suppression ratio. The enhanced modal loss difference, however, qualitatively describes that the new design can make the fundamental mode more dominant at a high current level than the conventional one. Another advantage of this structure is that modal loss can be controlled by changing the thickness of the absorption layer inside the cavity. The thickness of the layer can be controlled during crystal growth with great precision. In the calculation, a 40 nm GaAs layer is used as an absorption layer and it is placed near the first nodal point from the active layer.

## 4. Summary

A very simple and efficient evaluation procedure is suggested for the design of high power single mode VCSELs by reviewing the physical mechanisms that governs mode transition and simplifying the computation steps. In addition, the new structures are proposed and tested following the suggested evaluation tool. As a result, the new design exhibits much better stability of the fundamental mode over a wide current range than the conventional one.

**Acknowledgements**

This research was supported by the Kyungpook National University Research Fund, 2004. I would like to acknowledge Dr. Lee, Myung-Lae of the Electronics and Telecommunication Research Institute for his help in the calculation using ANSYS.